\documentclass[10pt, conference, compsocconf]{IEEEtran}

\usepackage{graphicx}
\usepackage{subfigure}
\usepackage{url}
\usepackage{xspace,colortbl}

\title{The ``Hot Potato'' Case: Challenges in Multiplayer Pervasive Games Based on Ad hoc Mobile Sensor Networks and the Experimental Evaluation of a Prototype Game}

\begin{document}

\author{\IEEEauthorblockN{Ioannis Chatzigiannakis, Georgios Mylonas, Orestis Akribopoulos, \\
Marios Logaras, Panagiotis Kokkinos, Paul Spirakis}
\IEEEauthorblockA{Research Academic Computer Technology Institute (RACTI) and\\
       Computer Engineering and Informatics Department, University of Patras, Greece\\
       Email: \{ichatz, mylonasg, akribopo, logaras, kokkinop, basilakn, spirakis\}@cti.gr}
}

\maketitle

\begin{abstract}
In this work, we discuss multiplayer pervasive games that rely on the use of ad hoc mobile sensor networks. The unique feature in such games is that players interact with each other and their surrounding environment by using movement and presence as a means of performing game-related actions, utilizing sensor devices.  We discuss the fundamental issues and challenges related to these type of games and the scenarios associated with them. We also present and evaluate an example of such a game, called the ``Hot Potato'', developed using the Sun SPOT hardware platform. We provide a set of experimental results, so as to both evaluate our implementation and also to identify issues that arise in pervasive games which utilize sensor network nodes, which show that there is great potential in this type of games.
\end{abstract}

\section{Introduction - Motivation}
\label{sec:intro}


The last few years we are witnessing a rapidly growing trend toward the interconnection of the digital and physical domains, as is also  evidenced by the outstanding activity in the wireless sensor networking research area and the continuous integration of sensing devices in multiple application domains. Even though we are using sensors in an ever-increasing multitude of ways, we have only scratched the surface regarding the use of sensors in entertainment-related applications. \textit{Gaming} is an important application domain that has not yet received a lot of attention by the wireless sensor networking (WSN) community. Although many different applications have been proposed for WSN, very few are related to mobile, interactive, multi-player games where users carry devices with sensing capabilities in order to interact with their environment and with each other.

Games of course have been a major part of the computer industry for the last decades, and are generally recognized as a means of pushing the technological boundaries, both in hardware and software. Recent advancements in mobile phones technology have introduced new products that integrate various kinds of sensors into the handsets. Given that in 2008 the total number of mobile phone subscribers has well surpassed the number of 3 billion, there exists a massive candidate user base for using such devices to play games with. It is our belief that there is great potential in combining sensors and mobile devices to produce new exciting entertainment applications. Wireless sensor networking is a field well-studied both in its theoretical and more practical aspects, and we feel that such knowledge can be utilized to provide new services and products. The use of sensors such as accelerometers, e.g., in the case of the Nintendo Wii gaming console, has already been proved a major success. There is an additional trend of detaching from traditional gaming environments, evident by the massive success of mobile platforms, like the Sony PSP and the Nintendo DS, or even mobile phones like the iPhone, that have also networking capabilities (as opposed to previous mobile gaming platforms). 

Lately, \textit{pervasive gaming} appeared as a hot topic, even though it is often used as a buzzword, meaning different things to different people, especially when discussed in the context of research work, and it is not easy to clearly define it as a gaming genre. Some examples of games that are usually placed in the pervasive gaming genre are alternative reality games, geotagging games, ubiquitous games, mixed reality games, urban games, etc. In this work we are interested in identifying the main issues that arise when combining pervasive gaming and WSN and to provide solutions and some preliminary results for some of them. Such games are largely based on three features, \textit{movement}, \textit{presence},\textit{ and other sensing inputs}, all provided by the use of sensor networking techniques. In this work, we focus on the use of only the first two features to demonstrate our point.

We believe that the knowledge stemming from the recent WSN-related research can be applied in the gaming domain so as to produce efficient systems. To make our point clear, we have implemented a game that falls into the category which we are discussing here, the ``Hot Potato'' game. We discuss the implementation approach we followed and highlight some key design issues. We utilize Sun SPOT \cite{SUNSPOT} nodes as our prototype implementation hardware platform, which provide the basic functionalities of wireless sensor network nodes. A number of services are currently implemented, allowing location awareness of wireless devices in indoor environments, performing sensing tasks while on the move, coordinating basic distributed operations (e.g., mutual agreement). The main contribution regarding the implemented game, is the provision of a series of experimental results, that were conducted with groups of human players, in order to validate the appeal of the game to the players and the overall suitability of the implementation. Our results indicate a very positive response from the people that participated and that the selected implementation platform is adequate in most of the aspects considered.

The remainder of the paper is organized as follows: in Section \ref{sec:related-work} we report on previous work and in the subsequent section we identify fundamental issues related to the application domain discussed in this work.  In Section \ref{sec:game-description} we describe the ``Hot Potato'' game and some of its implementation aspects. A set of experimental results are presented and discussed in Section \ref{sec:evaluation}, and we conclude our work in Section \ref{sec:conclusions}.

\begin{figure*}
\begin{center}
\begin{minipage}{0.31\textwidth}
        \centering
        \includegraphics[width=0.8\columnwidth]{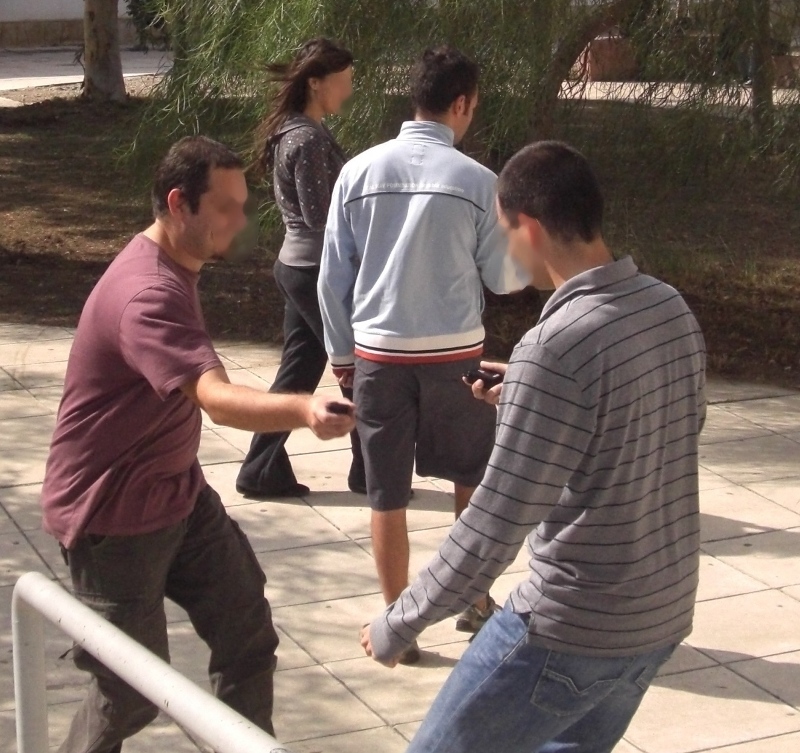}
\end{minipage}
\begin{minipage}{0.31\textwidth}
        \centering
        \includegraphics[width=1.1\columnwidth]{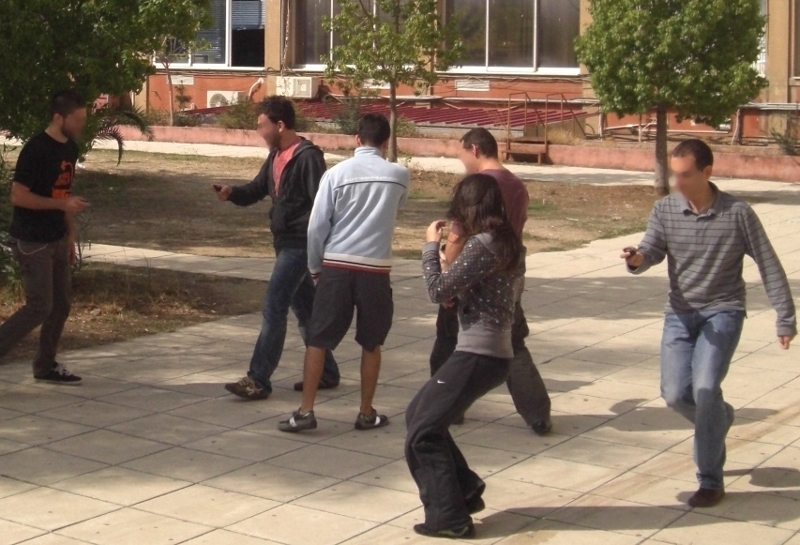}
\end{minipage}
\begin{minipage}{0.31\textwidth}
        \centering
        \includegraphics[width=0.75\columnwidth]{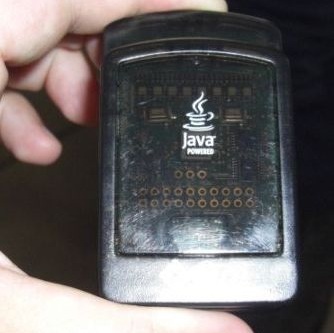}
\end{minipage}
\label{fig:hot-potato-gameplay}
\caption{Actual gameplay instances of the ``Hot Potato'' game using Sun SPOT enabled devices}
\end{center}
\end{figure*}

\section{Related Work}
\label{sec:related-work}

There is a large body of work regarding the pervasive games genre. The aim of the \textsf{IPerG} EU-funded project~\cite{IPERG} was the investigation of the pervasive gaming experience and the implementation of a series of ``showcase'' pervasive games. Several other works present implementations of pervasive games, see e.g., \cite{PIRATES, NETATTACK}. Some examples of well-known pervasive games are \textit{CatchBob!}, \textit{Uncle Roy All Around You}, \textit{SupaFly}~\cite{SUPAFLY}, \textit{Human Pacman}~\cite{human-pacman}. In \cite{atomic-actions} several interesting issues are raised, regarding the theory around pervasive games. In \cite{SUPAFLY} the authors evaluated how people perceive and play a pervasive game in normal, everyday settings. In general, most works focus on the design issues raised by specific games; some of these works additionally try to generalize such issues regarding the design of pervasive games overall, \cite{BENFORD05} and \cite{mythical} or provide surveys of existing approaches. In \cite{TRINTA08} the authors present scenarios that show the intended characteristics of pervasive multiplayer games and propose services for the development and deployment of crossmedia games, i.e., games that are played on multiple platforms with varying features.

The last couple of years we have witnessed the use of devices that utilize physical body movement input on a commercial scale, such as the Nintendo Wii. The big commercial success of this product has also led other major companies, such as Microsoft and Sony, to introduce their own versions of physical input devices. Such configurations range from using solely accelerometers and gyroscopes (Nintendo) to using a combination of such sensors along with cameras (Sony), and to using only cameras (Microsoft). Although these products allow for the detection of physical movement of the players, they differ somehow from our own platform and target application domain. We are investigating here applications and installations that: i) require input from multiple sources, and not only movement detection, but also (potentially) light, humidity, presence, temperature, etc., ii) are played by multiple players in non-controlled environments, so the use of cameras would potentially be non-applicable since players' views from cameras could be hindered by one another, iii) may require precise synchronization \textit{between} the people participating (in terms of time or movement), which also may be difficult to implement otherwise, iv) overall we aim at a different application domain. Finally, our work is based on open-source platforms and tools, whereas these are closed source commercial products, both hardware and software-wise.

There is also a certain body of work on using pervasive methods and tools for interactive installations with a certain museum/exhibition aspect, e.g., \cite{museum-ace-1, museum-ace-4, museum-ace-5}. Such works differ from our vision by not using multimodal sensor inputs in order to provide additional interaction with the users, they follow a centralized architecture and also do not provide much regarding services of synchronization and situation awareness. A quite thorough discussion on such matters, regarding the integration of sensor networks in entertainment-oriented applications, is included in \cite{MANI-EXPRESSION}.

Although there have been some attempts to develop multi-player games that rely on devices sensing the real world, these works are rather limited in number and scope, and are even less in a pervasive multiplayer context. Examples of WSN-based games are \cite{SENSORIUM,MOTTOLA}. In particular, \cite{MOTTOLA} describes a concept close to our own, implemented using a mobile agents middleware, but with a narrower overall scope and without the innovative user interface used in our work. Also, in \cite{BALLAGAS08} a tourism-oriented locative game is presented, which uses certain simple gestures to recognize user input, quite similar to the ones we are using, with the aid of mobile phone-integrated accelerometers, whereas we are currently using WSN nodes. A platform with different purpose than ours, but in the same spirit regarding the actual hardware platform and the user interface provided is described in \cite{SIFTABLES}. 
Regarding the networking technology used in similar games, \cite{KLONOWSKI04} describes the experiences from using Bluetooth to develop a multiplayer game. Our work currently focuses on using IEEE 802.15.4-compatible devices. To the best of our knowledge, our work is the first, in this context, that uses these type of devices. There is also a body of work in gesture recognition in resource-constrained devices, like the ones we envision using, such as in \cite{UWAVE}.

All of the works mentioned above differ from our vision, at least in the sense of their goals; we envision games that \textit{involve multiple players, rapid physical activity, gesturing}, whereas in the majority of the existing approaches, intense physical activity is not a prerequisite. Furthermore, we use small-in-size devices that are \textit{easy to carry and use}, whereas in the examples mentioned above PDAs or rather expensive mobile phones are used. This  limits the ways devices are used by the players, partly because of their size and the fear of potentially damaging them during gameplay. Also, in such games some portion of storytelling is involved in the process of making and actually playing the game; we currently lean toward \textit{less storytelling-based games}, which makes the development and set up of the games far less tedious. We think that these features will be critical for the wider acceptance of such games in the near future.

\section{Research Challenges}
\label{sec:challenges}

We make an attempt to identify the differentiating factors of our approach from already existing ones, along with some of the respective implementation requirements. 

\smallskip
\noindent\textbf{Simultaneous participation of multiple users:} we envisage games where groups of players participate, potentially in large numbers. The players will be in close proximity, in indoor or outdoor environments, and will have to engage in such applications by either interacting between themselves or with an infrastructure provided by the organizing authority. Depending on the nature of the game, players may have to cooperate or compete with each other, e.g., to reach the goals of a team-based game inside a museum, and this may be done in a real-time fashion. Regarding implementation, this assumes that there is a reliable neighborhood discovery mechanism, along with proximity detection, location-aware and context-aware providing mechanisms to the software and the players. These mechanisms are required to scale to a large number of players and to different area sizes.

\smallskip
\noindent\textbf{Multiple types of inputs:} we envisage the utilization of a plethora of inputs, the most general of which are presence, motion and other types of sensors. Such inputs may be provided by the mobile devices carried by the participating players or by the supporting infrastructure. For example, pupils or museum visitors will carry mobile devices that are able to sense their location (absolute or relative to each other and specific landmarks), their movement (both in terms motion detection and gesture recognition) and other physical measures (e.g., the device could sense if the player is in a warm/cold or light/dark place). Therefore an expandable architecture is required to cover all the different sensors that can be used on a single device and be reported to the upper layers of the system, along with mechanisms for reliable motion detection and gesture recognition. In the additional case of using cameras throughout the system, respective mechanisms for the same actions must be used.

\smallskip
\noindent\textbf{Non-conventional interfacing methods and use of actuators/haptics:} the players should be able to decipher both their personal and/or their team's status/score while engaging in the proposed interactive schemes, and also the system interfaces should reflect the location and context awareness inherent in such situations. The use of actuators such as operating lights and doors, haptic interfaces, etc., will enable a more immersive experience. 

\smallskip
\noindent\textbf{Distributed network operation:} the use of embedded sensors and ad hoc networking capabilities requires that the software executed on the mobile devices is based on lightweight mechanisms. The complex parts of the system's logic need to be implemented at the fixed infrastructure. Further functionalities may be required that rely on real-time coordination and complete knowledge of the users' whereabouts, or are executed in a disconnected part of the network. The system must be able to detect situations where real-time direct communication is not possible and activate delay-tolerant mechanisms to ensure the correct operation of the system and/or reliable multihop or multicasting mechanisms to cover all possibilities of communication between players and the infrastructure. It is therefore necessary for the architecture to be distributed, to involve a certain level of modularity and heterogeneity and support contingency plans that adapt the system performance to the actual conditions.

\smallskip
\noindent\textbf{Need for synchronization and coordination between players:} in most games players are competing or cooperating in order to reach/fulfill the goals set in a specific application. Players have to directly interact with each other and the overall system in a synchronized way. Such synchronization schemes should cover updates of the state of the players and the system, and also possibly coordinate the ways that the users move and act inside the playing field. Mutual exclusion, agreement and leader election mechanisms may be used to ensure the correct operation of the system.

\begin{figure*}
\centering
\includegraphics[width=0.78\textwidth]{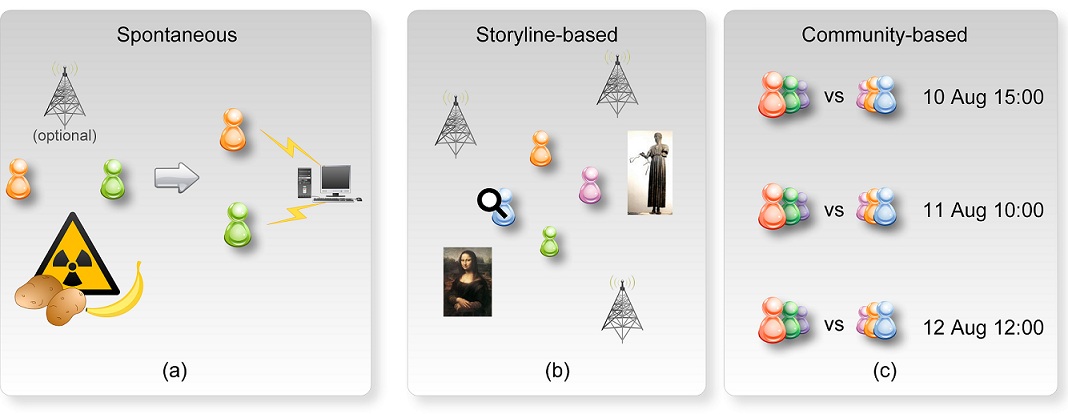}
\caption{\label{fig:categorization} (a) Spontaneous (hot potato/exploding bananas), (b) Storyline-based (museum excursion), (c) Community-based (augmented hide-and-seek)}
\end{figure*}

\smallskip
\noindent\textbf{Need for reliability and dependability:} since this type of games are to be played using even low-cost devices and in ``harsh'' settings, i.e., devices could be rebooted by mistake, accident during gameplay, etc., players need to be reassured that such characteristics will not interfere with the smooth operation of the games. The bookkeeping of activity and state during gameplay may be needed in certain games to ensure their correct operation, along with the provision of fault-tolerance mechanisms in order to recover from transient or permanent failures.

\smallskip
\noindent\textbf{The world-as-a-gameboard:} the games should be implemented in such a manner, so as to provide both the distributed nature and the context-awareness factors to the players. This is of course a central concept and challenge in pervasive gaming overall. How should the game handle the infrastructure islands, the periods when players are not inside them or the handovers from one to another?

\smallskip
\noindent\textbf{Multiple platforms / Crossmedia applications:} apart from using multiple, different inputs, this kind of applications might else be based on a variety of platforms, depending on the environment they are played at (e.g., play the game in office on weekdays or at home on weekends) or even the logical level involved at separate time instances (e.g., playing the game with a mobile device and then playing other aspects of the game or watching statistics on a desktop computer). The diversity in inputs, user interfaces and capabilities makes it very complex to provide a unified experience.

To further emphasize the diverse possibilities of the projected applications, we categorize them in terms of interaction between users, existence of an infrastructure and time-wise operation:

\begin{itemize}
	
	\item \textit{Spontaneous}: games are setup instantly between players, e.g., when a having a break from work at the office/school. No specific infrastructure is needed, with communication and sensing inputs be provided only by the mobile devices the players are carrying themselves. Game results are uploaded to a central entity after the game is over, in order to gather statistics or gain/gather game score points. No limits are imposed whatsoever regarding time and place of the gamefields.
	
	\item \textit{Storyline-based}: a degree of infrastructure is used in order to provide presence and other sensing input combined with a mixture of a game scenario and timely events. A central authority coordinates the game as mandated by a given scenario. An example of such a game, could be an entertainment/educative installation inside a museum, where players are given a starting scenario and must discover elements to move forward in the game by visiting places, etc. The infrastructure intervenes both in the spatial and temporal domains to define the possible game outcomes, but all is performed within specifically defined limits.
	
	
	\item \textit{Community-based}: this is basically the re-hash of traditional games, like ``hide and seek'', augmented with the aid computing devices and sensing abilities, which can help in game operation, rules arbitration, statistics book-keeping, etc. They can be both infrastructure-less or infrastructure-based, but the distinct element here is the concept of the gaming community. This means, that there is a community playing games of various sorts, with the playing activity spanning across a long time period and with large gaming interaction among the members of the community.
\end{itemize}

This categorization is also depicted in figure \ref{fig:categorization}.

\section{The Hot Potato Game}
\label{sec:game-description}

We now present a simple game of the application type we are discussing in this paper. We discuss some of the inner works of our implementation, in order to give answers  and further insight into the aforementioned challenges.

The game is played by a group of players gathered at a specific place. Each player holds a device that has the potential of generating a ``Hot Potato'' after the game begins. Each such hot potato has a per-second decreasing counter that ticks for a certain period of time until it goes ``boom''. Players try to ``survive'' by passing the potato on to another player, thus eliminating the danger of the potato ``exploding'' on their device. However, the time counter value is preserved between potato passings. When, inevitably, a potato blows, the respective player who was the last carrying the potato is disqualified, decreasing the number of players attending the game. The winner of the game is the player standing alive while all opponents are eliminated.

The players interact with each other by using movement and presence as a means of performing game-related actions. To pass the potato the player needs first to approach an opponent by moving close to her. Then the player passes the potato to the nearby player by violently moving her device to the right/left direction (i.e. making a game action/gesture). 

The players interaction also affects the generation of new potatoes. Whenever some players choose to be close to each other (i.e., close proximity) the probability of generating a new potato is reduced substantially. Therefore it is mutually beneficial for players to stay close together. Still, eventually at some point one of them will get a new potato. Now, the strategy changes since being close to the player holding the potato makes it easier for him to pass the potato.

\smallskip
\noindent\textbf{Hardware/software Platform:} We implement our game using Sun's SPOT platform \cite{SUNSPOT}, as the hardware platform for the players' devices. It is a small, battery-operated device running the Squawk Java Virtual Machine, which acts as both an operating system and a software application platform, allowing programming of the devices in the Java Micro Edition (J2ME) platform. It uses an $180$MHz ARM 9 processor with $512$KB of RAM and $4$MB Flash. An IEEE 802.15.4 compliant CC2420 Chipcon transceiver is used for communication. They also include a very simple user interface ($2$ buttons and $8$ LEDs) and a number of sensors (accelerometer, thermistor, light). 

We decided to use the particular platform due to the available computational resources; other WSN platforms usually offer $10$KB of RAM and a processor at $8$MHz. The fast ARM processor and the rich RAM allows more relaxed programming. The provision of $8$ LEDs and $2$ buttons is also significantly better than the other platforms that usually provide 2-3 LEDs and at most 1 button. The additional LEDs \textit{significantly improve the interfacing methods} of the device. Furthermore developing applications for JAVA Mobile Edition \textit{improves the portability} of our code to almost all types of mobile phones and portable devices. 

\smallskip
\noindent\textbf{Input mechanisms for player interaction:} Player gestures to pass the potato are detected using the internal accelerometer. We aimed at user-independent gesture recognition, with no training phase involved. Our code can recognize four basic gestures: i) clockwise, ii) counter-clockwise and iii) violent movement with direction to the right/left. These $4$ gestures are very simple and are in fact a subset of the $8$ gestures described in \cite{NOKIA-ACCELEROMETER}, that were deemed as most appropriate for home automation. 

The interactions based on players proximity are implementation using the IEEE 802.15.4 radio. In contrast to Bluetooth, devices can exchange data without the need to establish a connection, does not enforce a master/slave relation and supports concurrent transmissions of messages to multiple devices. The most important characteristic is the ability to reliably discover neighboring devices within less than $100$ms time period while Bluetooth devices may take up to $10.24$sec \cite{DBLP:conf/loca/HayH09}. This \textit{significantly improves proximity detection} capabilities among players. 

The \textit{Echo Protocol} is a core service of our implementation that enables neighbor discovery and enables connectivity. Through the Echo Protocol, a player device discovers other neighboring devices and observes their connectivity status. 
Devices form a network characterized by high mobility and variable transmission ranges. The topology of such network changes dynamically in an unpredictable way, as players move without following any predefined pattern. 

The Echo Protocol is designed to run on resource-limited devices. It is robust, able to adjust quickly to frequent and significant topology changes and capable of distinguishing the different roles of the discovered neighboring nodes (i.e., player, infrastructure). In addition, it allows customization of propagated messages and most importantly it is able to recognize whether the communication with the surrounding devices is bi-directional or not.


\smallskip
\noindent\textbf{Coordination of player interactions:} In general, players interact in pairs or in groups by executing (simultaneously or not) \textit{actions}. For each game there is a set of well defined actions that the players can perform. It is important that the actions are performed efficiently with near-real time response and that fault tolerance mechanisms exist for protecting against communication failures, common due to the wireless medium used. Moreover, the existence of a reliable mechanism, which will prevent malicious player behavior and apply different game rules is absolutely essential. For these reasons, we implemented the \textit{Action Protocol}; this is a two-phase commit protocol, which supports different types of actions, allows more than two devices to interact with each other and reduces the risk of unexpected or adversarial behavior (e.g., cheating). It guarantees that a potato will not be lost due to communication failure (e.g., messages lost during dense player interactions).

\smallskip
\noindent\textbf{Towards reliability and multi-games support:} During the evolution of the game, player devices may fail permanently due to power loss (devices are battery operated) or temporarily due to an internal error that occurred at the virtual machine. In other cases devices may reboot either by mistake or maliciously. To avoid such situations that may damage the player experience we implement the \textit{Storage module} that is responsible for providing persistent storage capabilities. This module stores the state of the device and the various game events created on the $4$MB flash. In the case the failure is transient, the data are retrieved during the reboot of the device; the countdown timer of an active potato will continue (with a small delay of $3$sec needed for the device to reboot). In case of a permanent failure, the player may still recover the stored events by plugging the device to an external power source or by attaching it to a PC.

The storage module is also responsible for storing all game events when the player decides to play another game yet she wishes to keep track of the old game statistics. When all the Hot potato games are concluded she may return to her PC and upload the game data to a higher layer application (e.g. community-based application). This is done by simply pressing one of the buttons that prepare the device for another game.

\smallskip
\noindent\textbf{Support for Storyline-based and Community-based extensions:} An interesting feature of our implementation is that external platforms may interface player devices in real time during the evolution of a game. This is done by installing an infrastructure of one or more \textit{Stations} at the area where the games are conducted. These Stations may be laptops, PDAs or SPOT devices; they can be fixed or mobile and form a backbone network. During the course of the game the data generated by the player devices and game statistics are communicated to nearby Stations. One particular Station plays the role of the \textit{Engine}, responsible for the central coordination of the Stations and generally of the game (e.g., checking the rules of the game). The Engine is also responsible for communicating the progress of the game to external platforms and upper layer applications. The Station and Engine can execute their tasks asynchronously in a pipelined way. This way a Station sends information regarding the realization of a number of game events to the Engine, without waiting for the Engine to complete forwarding all these data to the upper software layers. When this feature is enabled the storage module is no longer needed since game data are uploaded to the higher level application during the game.

\smallskip
\noindent\textbf{Delay-Tolerant Service:} Due to various reasons (e.g., arbitrary movement of players, game strategies) communication with the ``backbone'' infrastructure may not be always possible. During this period the evolution of the game should not be affected, as players interact with each others and create events. For this reason we have created, the \textit{Delay-Tolerant Service} (DTS) that allows operation on both connected and disconnected modes. When communication with the infrastructure is available, data are transmitted (either directly or in multihop fashion), while when this is not possible, data are stored and sent afterwards - when communication is establish. Thus, players can enter and leave from the range of the infrastructure, enjoying the games without any limitations. In any case, however, communication with the infrastructure has to be established, even at a later point in time (e.g., after the game finishes), in order to keep the history of played games consistent.

\section{Evaluation of the Game}
\label{sec:evaluation}

%

We chose to put our system under test in real life conditions - it is our opinion that the fundamental evaluation of a game should be done while groups of human players are actually playing it. We chose to conduct experiments both indoors and outdoors, since our vision is that this kind of games will be played in both environments. More specifically, the outdoor experiment was conducted in the backyard of our research institute, an open area covered with grass and with no obstacles (trees, bushes, etc.) inside the field. The indoor experiments were conducted in i) a room measuring $10$ meters by $15$ meters, with a height of $3$ meters, and ii) an atrium measuring $15$ meters by $15$ meters and a height of $8$ meters, with $4$ pillars ($1$ meter diameter) close to its center. Both rooms/playfields had concrete walls. 

Regarding the experiments, game sessions were organized in groups of $5 \ldots 14$ people. The total number of the people that took part in the experiments was $23$.
All players were aged between $23$ and $37$, with the $75\%$ of them having an engineering background and the rest coming from other disciplines. Each group of players participated in multiple game sessions, with some participating in multiple sessions with different number of players. Regarding the quantitative measures of the experiments, additional software modules were introduced to measure a variety of metrics.

\subsection{Player survey results}

At the end of the experiments, all of the participants (\textit{game developers did not participate}), were given a questionnaire to fill out. The questions posed to the players ranged from opinions about the overall satisfaction/fun they had from participating to the organized game sessions, to questions about how they would react to enhancements to the hardware platform and the user interface.

From the results, the following basic points are evident:

\begin{itemize}
	
	\item All players were positive about the fun factor of the game and the ease of use of the provided device; $24\%$ were strongly positive and $71\%$ positive about the appeal of the game to them.
	
	\item The majority of  players were positive about the physical activity involved in the game ($82\%$ positive or strongly positive and $18\%$ neutral), with the vast majority ($16$ out of $17$) also stating that the use of gestures is more fun than using buttons.
	
	\item The response was more neutral regarding the current gesture recognition implementation, with $24\%$ being positive, $42\%$ being neutral and $35\%$ being negative or strongly negative, probably due to the user-independent recognition strategy we chose.
	
	\item The game is more fun when there are more players involved, with $76\%$ of the participants being strongly positive about large player groups.
	
	\item Players said that additions in the user interface, such as a screen and vibration would be welcome, even when not coming from a CS background. The reaction to sound effects addition was much more neutral ($36\%$ neutral or negative).
	
	\item The overall operation of the system was perceived by the users as satisfactory in both indoor and outdoor environments, with the reaction to the indoor case being a little more positive.
	
\end{itemize}

\subsection{Performance evaluation results}

Following the logic of our general software architecture, we have incorporated a number of software modules implementing the observer software pattern in order to record all characteristics of interest for the experiments. E.g., every time a gesture was performed and recognized correctly by the device, an event was created that informed the experiments observer about this specific action. We have limited the number of recorded features in order to minimize the computational overhead. All recorded data were kept in memory during the gameplay (since writing to the device's flash memory incurred significant overhead) and after the respective game session was over, it was transmitted back to a base station. All features recorded were the same in all devices, except from the CPU load logging, which was recorded in only two devices for each session, since the Squawk VM does not provide any facilities to log such data, and we wanted to avoid adding extra overhead to all of the devices at the same time.

\begin{figure}
\centering
\includegraphics[width=\columnwidth]{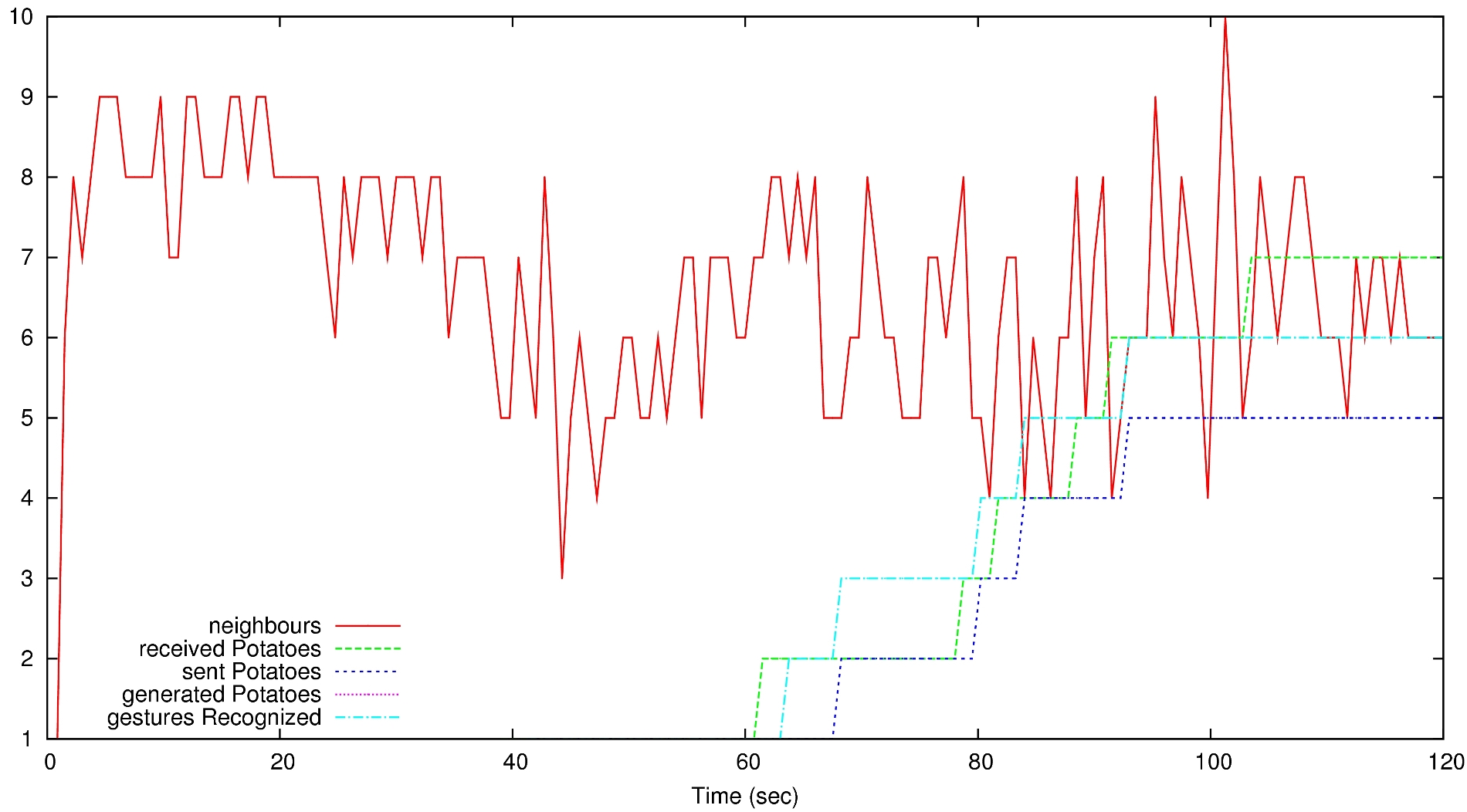}
\caption{\label{exp:1} Players, potatoes and interactions}
\end{figure}

\begin{figure}
\centering
\includegraphics[width=\columnwidth]{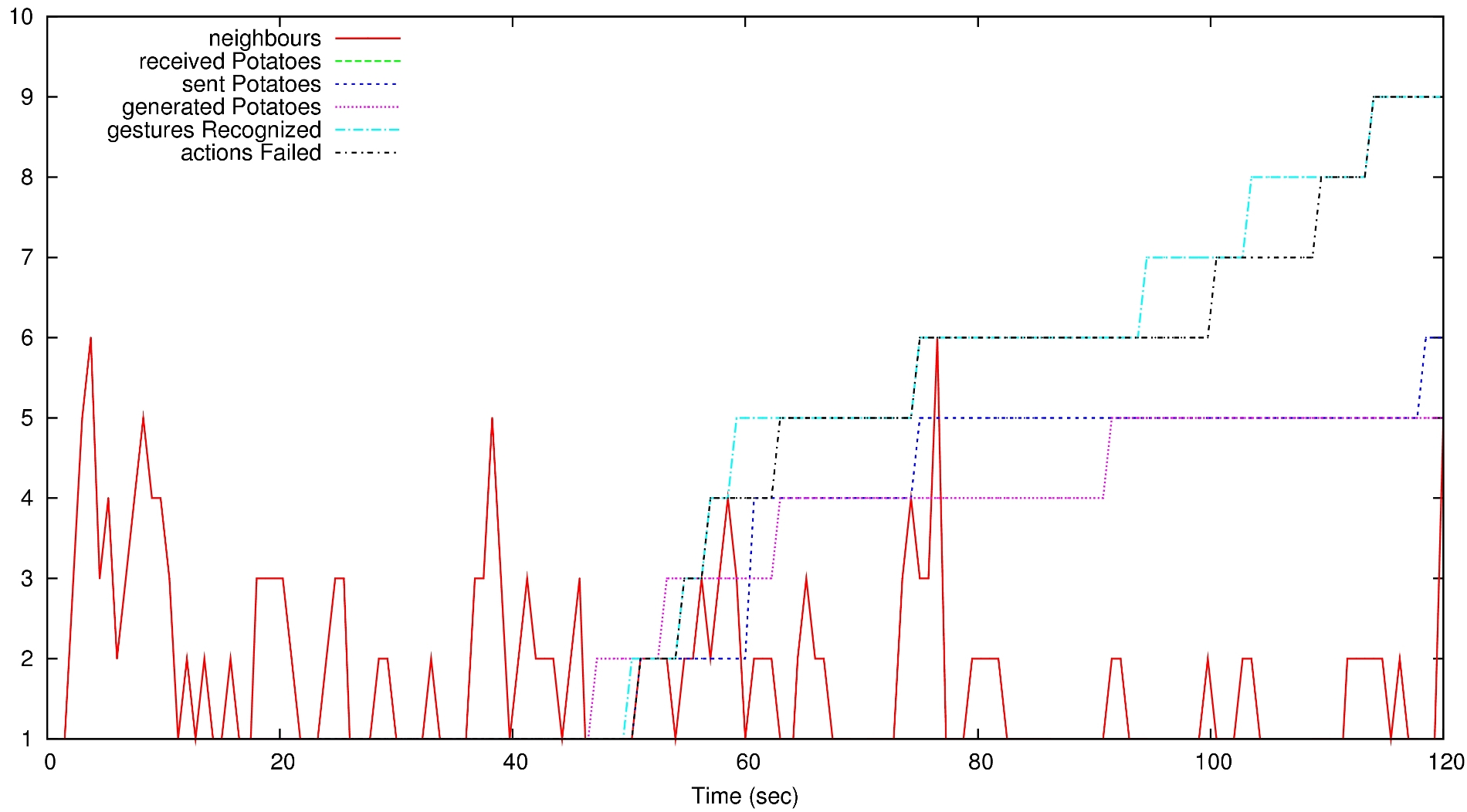}
\caption{\label{exp:1b} Players, potatoes and interactions}
\end{figure}

More specifically, we monitored the following metrics; they were chosen so as to measure both the efficiency of the implementation and the overall fluidity of the gameplay:

\begin{itemize}
	
	\item \textit{Neighbors}: the number of bidirectional links each device has established in each moment in time.
	
	\item \textit{Potatoes received/sent}: the number of ``potatoes'' each player has received/sent during gameplay.
	
	\item \textit{Potatoes generated}: the number of ``potatoes'' generated by each player's device.
		
	\item \textit{Gestures recognized}: the number of gestures that were successfully recognized.
		
	\item \textit{Failed actions}: the number of times when players tried to pass a ``potato'' but failed, although the gesture was recognized by their own device.	
%
	
	\item \textit{CPU usage}: an estimate of the load during gameplay.
	
	\item \textit{Memory usage}: the memory usage of the virtual machine after loading the Hot Potato midlet.
	
\end{itemize}

We do not present here any energy consumption results, due to the fact that the duration of a ``hot potato'' game is very short ($2-3$ minutes) and that we have not tested the game for extended time periods, i.e., hours of continuous gameplay. Our measurements indicate that energy consumption is quite minimal during such game sessions and does not have a big impact on the device's battery.

Figure~\ref{exp:1} depicts the dynamicity of the network for a game that lasted about $2$ minutes. At some points, the device neighborhood changed from $9$ neighbors to $3$ in less than $3$ sec. This clearly shows \textit{the high dynamics of the games we envisage}. The experiments show that the choice of the IEEE 802.15.4 radio module in this case is valid, since it is capable of detecting the proximity of the other players fast enough to support the game rules. 

In terms of game play and game strategies, the increased activity of the player seems to pay off for at least $1$ min. After that point, she ends up receiving a potato from a fellow player; within $2$ sec, realizing she now carries a hot potato, she makes a gesture to get rid of the potato, that is recognized by the device and the \textit{action protocol} is activated. It requires about $3$sec to negotiate the passing of the potato on to one of her opponents. This fast reaction saves her about $10$ sec of gameplay and then a $10$ sec period starts of continuous potato passing and subsequent running and chasing. Overall, the potato passes $4$ times. Eventually, she ends up with a potato that ``explodes'' before she can pass it on to a nearby player, therefore ending the game for her. This scenario is a characteristic example of our vision regarding \textit{pervasive games that involve multiple players, rapid physical activity and gesturing}.

A different game strategy seems to be followed by the player whose device statistics are depicted in Figure~\ref{exp:1b}. This player tried to stay at the ``periphery'' of the board, encountering very few other players with the incentive of reducing the chances of getting a potato. This concept seems to be successful since he never received any potato throughout game. On the other hand, the reduced number of players within his neighborhood increased the changes of generating a new potato. After about $40$sec a new potato was eventually generated. Unfortunately, due to being at a distance from the other players, he could not find an opponent for a sufficiently long period of time to pass the potato. Although the player performed the corresponding gestures and the device successfully recognized them, due to the dynamicity and low density of the network the Action protocol failed to pass the potato. Even when he eventually succeeded, new potatos were repeatedly being generated until the player was eliminated. We believe that this example demonstrates the \textit{diversity of player strategies and the combinations of actions} that can be performed even in such a simple game.

\begin{figure}
\centering
\includegraphics[width=\columnwidth]{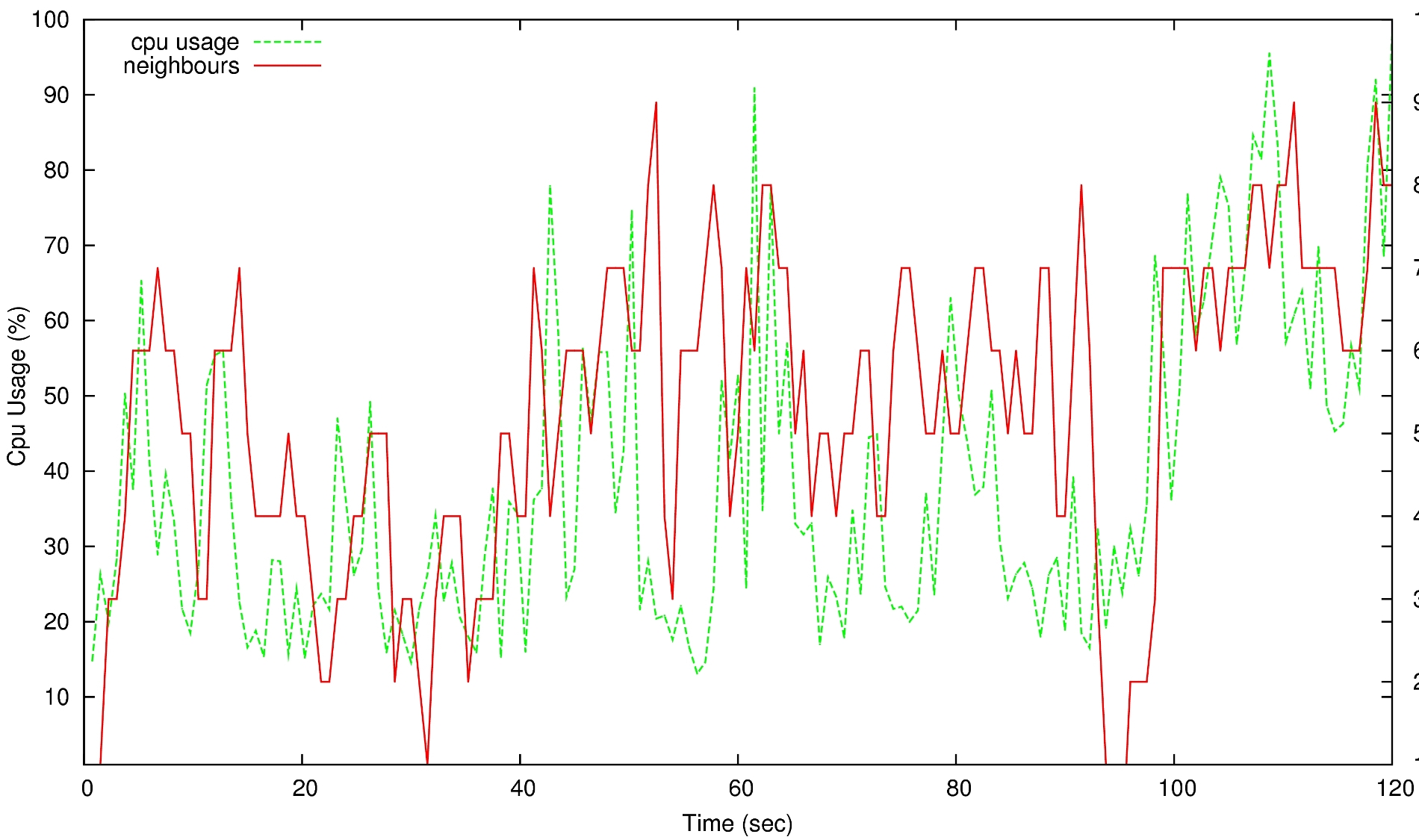}
\caption{\label{exp:2} Player density and CPU usage}
\end{figure}

\begin{figure}
\centering
\includegraphics[width=\columnwidth]{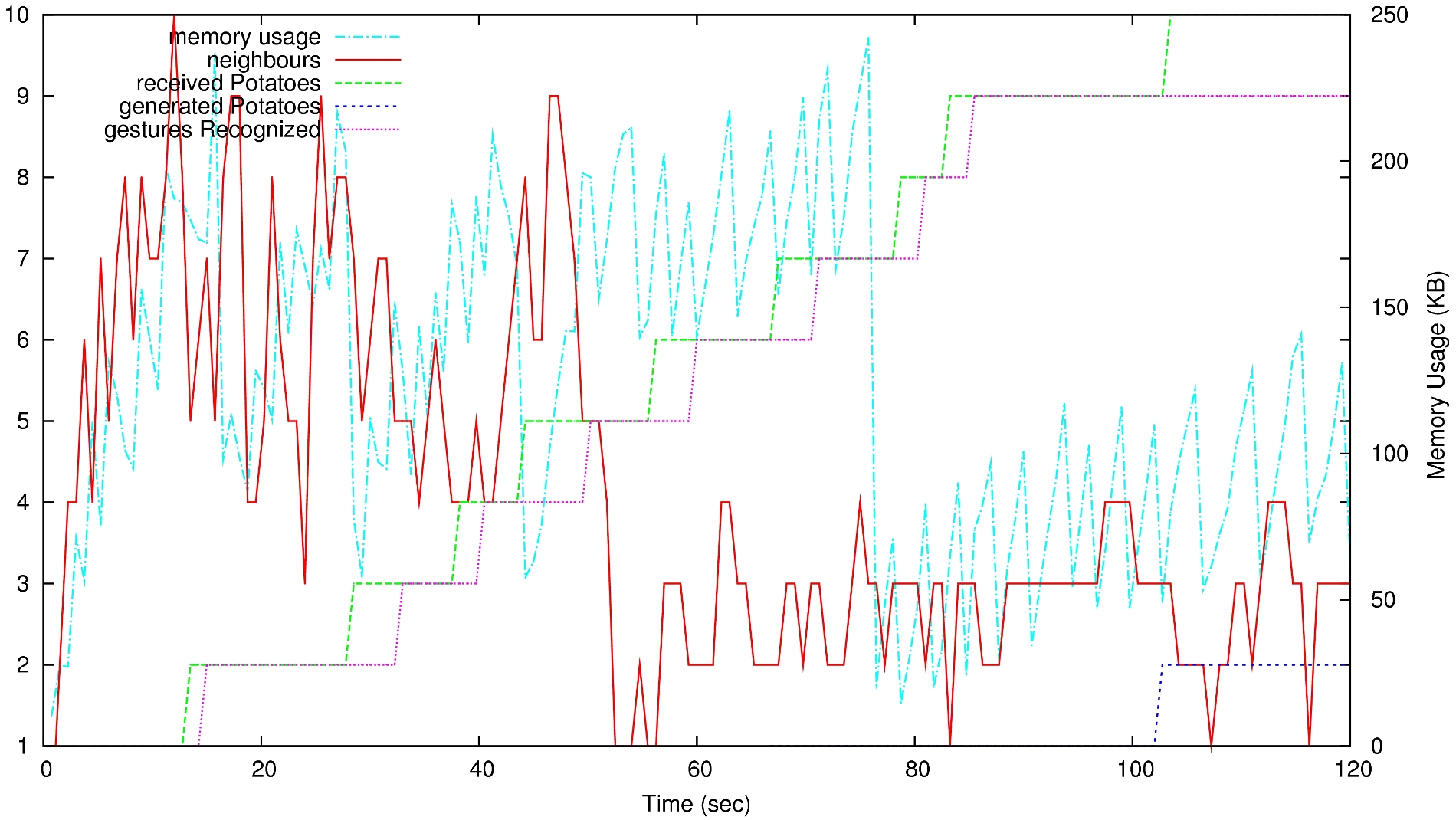}
\caption{\label{exp:3} Potatoes, Interactions and Memory usage}
\end{figure}

Figure~\ref{exp:2} depicts the CPU usage and the dynamicity of the network for a period for $2$ min. It helps us demonstrate the computational complexity of the \textit{Echo protocol} to detect the \textit{bidirectional neighbors} of the device. Although the majority of theoretical works consider wireless links to be symmetric, in practice most of the times this is not true for a multitude of reasons (e.g., orientation of antennas, different battery levels, obstacles, etc.). This fact significantly complicates game design; a unidirectional link cannot provide acknowledgements of receipt of messages. Of course, gesture recognition, LED operation, button control and garbage collection also requires CPU cycles, still, the \textit{majority of the CPU load is due to the Echo protocol}. The experiment indicates that  the particular ARM9 processor used in the game devices can be used for playing games with \textit{simultaneous participation of at least $14$ players} without hindering the player experience.

We also include Figure~\ref{exp:3} to examine the memory usage of the game. It seems that the $512$KB of memory sufficiently support the size of the Squawk VM, the code of the Hot Potato midlet and the entities generated during the gameplay. It is in the nature of Java to create a large number of object instances; the Echo protocol generates $1$ new object per nearby device every $500$ms, the Action protocol generates $3$ new objects per player action, etc. Twice during the evolution of the game the memory was exhausted and the garbage collector (GC) was required to clean up unused memory. Still, throughout the game evolution GC was invoked continuously to free up smaller amounts. This also increased the CPU load. It is evident that if we had chosen a different WSN platform, developing the Hot potato game would be very complicated, if not impossible.

\section{Conclusions - Future Work}
\label{sec:conclusions}

In this paper, we have discussed about a new category of pervasive games, that relies on the use of ad hoc mobile sensor networks for their operation. We have also presented a set of research challenges which we believe are the most important ones rising from this new vision. Moreover, we presented the rules of a prototype game we propose in order to demonstrate the potential of this new game category, and briefly discussed its current implementation, which relies on the use of the Sun SPOT sensor network devices. ``Hot Potato'' is an example of using non-conventional user interface methods to breathe new life into familiar concepts, like the multiplayer games played out in open space. Our implementation offers services for the completely distributed operation of the game and synchronization between players, enabling the instant setup of games at anyplace.

Through a series of experiments we evaluated our implementation and the response of human players to our game. Our first results indicate a very positive response and that the selected class of devices can sufficiently support this type of games. Up to $14$ players can participate in a game session simultaneously in a completely distributed environment; above this limit, inherent technology factors come into play and prevent a seamless gaming experience. Simple synchronization and coordination mechanisms such as the protocols discussed in section \ref{sec:game-description} can provide a sound basis for developing this new category of games. Players perceived the whole operation of the implemented game as satisfactory and welcomed the idea of being able to play in physical space, and even considered positively the idea of playing games with anyone, anyplace, anytime. Furthermore, the efficient implementation of synchronization and coordination between multiple groups of players simultaneously enables a whole new level of user interaction activities, as seen by the strongly positive perception of playing in large groups.

Our future work includes the implementation of additional games, the refinement of our implementation, especially with regard to gesture recognition, the conduct of game sessions in larger scale and possibly porting the game to other platforms.

\section{Acknowledgements}

This work has been partially supported by the European Union under contract numbers
IST-2005-15964 (AEOLUS) and ICT-2008-215270 (FRONTS). We also wish to thank I.~Mavrommati for her insightful comments and ideas.


\end{document}